\documentclass[11pt,a4paper,twoside]{article}
\usepackage{graphicx}
\usepackage{color}
\usepackage{epstopdf}
\usepackage{lscape}
\usepackage{amsmath}
\usepackage{hyperref}
\usepackage{setspace}
\begin{document}
\baselineskip=0.20in
\vspace{20mm}
\baselineskip=0.30in
{\bf \LARGE
\begin{center}
$\kappa$ state solutions for the fermionic massive spin-$\frac{1}{2}$ particles interacting with double ring-shaped Kratzer and oscillator potentials
\end{center}}
\vspace{2mm}
\begin{center}
{{\bf K. J. Oyewumi}}\footnote{\scriptsize E-mail:~ kjoyewumi66@unilorin.edu.ng}, {\bf B. J. Falaye}\footnote{\scriptsize  Corresponding author E-mail:~ fbjames11@physicist.net}, {\bf C. A. Onate}\footnote{\scriptsize E-mail:~ onateca12@gmx.us}, \\
\small
\vspace{2mm}
{\it Theoretical Physics Section, Department of Physics\\
 University of Ilorin,  P. M. B. 1515, Ilorin, Nigeria.}\\

\vspace{5mm}
{\bf O. J. Oluwadare}\footnote{\scriptsize E-mail:~ jostimint@yahoo.com}\\
\vspace{2mm}
\small
{\it Department of Physics, Federal University Oye-Ekiti,\\
P. M. B. 373, Ekiti State, Nigeria}

\vspace{5mm}
{{\bf  W. A. Yahya}}\footnote{\scriptsize E-mail:~ wazzy4real@yahoo.com}\\
\vspace{2mm}
\small
{\it College of Pure and Applied Sciences, Department of Physics and Material Science,\\
Kwara State University, P. M. B. 1530, Ilorin, Nigeria.}
\vspace{2mm}
\end{center}

\noindent
\begin{abstract}
\noindent
In recent years, an extensive survey on various wave equations of relativistic quantum mechanics with different types of potential interactions has been a line of great interest. In this regime, special attention has been given to the Dirac equation  because the spin-1/2 fermions represent the most frequent building blocks of the molecules and atoms. Motivated by the considerable interest in this equation and its relativistic symmetries (spin and pseudospin) in the presence of solvable potential model, we attempt to obtain the relativistic bound states solution of the Dirac equation with double ring-shaped Kratzer and oscillator potentials under the condition of spin and pseudospin symmetries. The solutions are reported for arbitrary quantum number in a compact form. the analytic bound state energy eigenvalues and the associated upper- and lower-spinor components of two Dirac particles have been found. Several typical numerical results of the relativistic eigenenergies have also been presented. We found that the existence or absence of the ring shaped potential potential has strong effects on the eigenstates of the Kratzer and oscillator particles with a wide band spectrum except for the pseudospin-oscillator particles where there exist a narrow band gap.
\end{abstract}


potential; double ring-shaped oscillator potentials.


\section{Introduction}
The concept of pseudospin was first introduced into nuclear physics in 1969 by Hecht and Adler \cite{HeA69} and was originally observed in spherical atomic nuclei, in the same year by Arima et al. \cite{R1}. This is to elucidate the experimental observation of the quasidegeneracy in single-nucleon doublets between normal parity orbitals $(n \ell,j = \ell +1/2)$ and $(n-1, \ell+2, j = \ell +3/2)$, where $n$, $\ell$ and $j$ denotes the radial, orbital, and total angular momentum quantum numbers, respectively. The total angular momentum is $j=\tilde{\ell}+\tilde{s}$, where $\tilde{\ell}=\ell+1$ is pseudo-angular momentum and $\tilde{s}$ is pseudospin angular momentum\cite{Gi97, Gi05} 

Within the framework of Dirac theory, it has been used to form an effective nuclear shell model \cite{HeA69, TrE95}. It has also be used to explain different phenomena in nuclear structure, for example the deformation, super-deformation, identical bands, and magnetic moment [6-10].
The spin symmetry is relevant for mesons \cite{PaE01}. The spin symmetry occurs when $\Delta(r)=V(r)-S(r)=$constant or the scalar potential $S(r)$ is nearly equal to the vector potential $V(r)$, i.e., $S(r)\approx V(r)$. On the other hand, the pseudospin symmetry occurs when $\Sigma(r)=S(r)+V(r)=$ constant or $S(r)\approx -V(r)$ [12-14]. 

In the recent years, there have been several worthy attempt in finding the solution of the Dirac equation with various potential under the condition of spin and pseudospin symmetry. The energy spectra and the corresponding two-component spinor wave functions of the Dirac equation for the Rosen-Morse potential with spin and pseudospin symmetry have been obtained by Oyewumi and Akoshile \cite{OyA12}. Falaye and Oyewumi \cite{FaO11} obtained the solutions of the Dirac equation with spin and pseudospin symmetry for the scalar and vector trigonometric scarf potential in arbitrary D-dimensions within the framework of an approximation scheme to the centrifugal barrier by using the Nikiforov-Uvarov method. Aydo$\acute{g}$du et al., by using the Nikiforov-Uvarov (NU) method, investigated the pseudospin and spin symmetric solutions of the Dirac equation for the scalar and vector Hulth$\acute{e}$n potentials with the Yukawa-type tensor for an arbitrary spin-orbit coupling quantum number \cite{AyE12}. 

Furthermore, some authors have investigated the spin symmetry and Pseudospin symmetry under the Dirac equation in the presence and absence of coulomb tensor interaction for some typical potentials such as the Harmonic oscillator potential [16-25], 
Coulomb potential \cite{HamE101, ZaE11}, Woods-Saxon potential \cite{GoS05, AyS10}, Morse potential [30-35],  
Eckart potential \cite{JiE06, ZhE08}, ring-shaped non-spherical harmonic oscillator \cite{GoE06}, P\"{o}schl-Teller potential [39-43], 
three parameter potential function as a diatomic molecule model \cite{JiE072}, Yukawa potential [45-49], 
pseudoharmonic potential \cite{AyS101}, Davidson potential \cite{SeH09}, Mie-type potential \cite{HamE102}, Deng-Fan potential \cite{MaE122}, hyperbolic potential \cite{HasE11} and Tietz potential \cite{HasE122}.
 
Recently, Castro addressed the behavior of the Dirac equation with scalar $(S)$, vector $(V)$ and tensor $(U)$ interactions under the $\gamma^5$ discrete chiral transformation. By using this transformation he obtained solutions for the Dirac equation with spin $\Delta=V-S=0$ and pseudospin $\sum=V+S=0$ symmetries, which includes a tensor interaction \cite{Ca12}. Ikhdair solved the Dirac equation with the screened Coulomb (Yukawa) potential for any arbitrary spin-orbit quantum number $\kappa$. He found that a wide range of permissible values for the spin symmetry constant $C_s$ from the valence energy spectrum of particle and also for pseudospin symmetry constant $C_{ps}$ from the hole energy spectrum of antiparticle \cite{Ikh12}.

The unbound solutions of generalized asymmetrical Hartmann potentials under the condition of the pseudospin symmetry have been presented by mapping the wave functions of bound states in the complex momentum plane via the continuation method \cite{GuE07}. Very recently, Guo explored the pseudospin symmetry by using the similarity renormalization group and shown explicitly the relativistic origin of the symmetry \cite{Guo12}. Chen and Guo \cite{ChG12} investigated the evolution of the spin and pseudospin symmetries from the relativistic to the nonrelativistic and explore the relativistic relevance of the symmetries. 

Zhou et al. \cite{ZhE03} used the relativistic mean field theory to investigate single anti-nucleon spectra. The spin symmetry in anti-nucleon spectra of a nucleus have been tested by investigating the relations between the Dirac wave functions of the spin doublets and examining these relations in realistic nuclei within the relativistic mean-field model \cite{HeE06}. From the Dirac equation, the mechanism behind the pseudospin symmetry was studied by Meng et al. \cite{MeE98, MeE99} and the pseudospin symmetry was shown to be connected with the competition between the centrifugal barrier  and the pseudospin orbital potential, which is mainly decided by the derivative of the difference between the scalar and vector potentials. 

Very recently, Song et al. \cite{SoE11} studied the effects of tensor coupling on the spin symmetry of $\bar{\Lambda}$ spectra in $\bar{\Lambda}$-nucleus systems with the relativistic mean-field theory. Shortly thereafter, Lu et al. \cite{LuE12} show that the conservation and the breaking of the pseudospin symmetry in resonant states and bound states share some similar properties. Furthermore in 2013, By examining the zeros of Jost functions corresponding to the small components of Dirac wave functions and phase shifts of continuum states, Lu et al. \cite{LuE12} also showed that the pseudospin symmetry in single particle resonant states in nuclei is conserved when the attractive scalar and repulsive vector potentials have the same magnitude but opposite sign \cite{LuE13}. Li et al. \cite{LiE13} proved that the spin-orbit interactions always play a role in favor of the pseudospin symmetry, and whether the pseudospin symmetry is improved or destroyed by the dynamical term relating the shape of the potential as well as the quantum numbers of the state.

A very recent work \cite{Guo12} filled the gap between the perturbation calculations and the supersymmetry descriptions by using the similarity renormalization group technique to transform the Dirac Hamiltonian into a diagonal form. Liang et al. explored the origin of pseudospin symmetry and its breaking mechanism  by combining supersymmetry quantum mechanics, perturbation theory, and the similarity renormalization group method \cite{R3}. By following the lead of authors in ref \cite{R3}, Shen et al. \cite{R4} further discuss
the spin-orbit effects on the pseudospin symmetry within the framework of supersymmetric quantum mechanics. By using the perturbation theory, the author demonstrate that the perturbative nature of PSS maintains when a substantial spin-orbit potential is included. Guo and co-researcher investigated the relativistic symmetry for an axially deformed nucleus by comparing the contributions of every term to the single particle energies and their correlations with the deformation to disclose the origin of the pseudospin symmetry and
its breaking mechanism in deformed nuclei \cite{R5}.

It was shown recently in ref \cite{R6}, that the experimental information on the energies of the $\gamma-$vibrational states in nuclei can be used to determine a splitting of the pseudospin doublet. Very recently, Alberto et al. \cite{R7} derived the node structure of the radial functions which are solutions of the Dirac equation with scalar and vector confining central potentials, in the conditions of exact spin or pseudospin symmetry. Recently, Ikhdair et al. \cite{R8} introduce a novel potential called as the generalized inverse quadratic Yukawa potential in the form of Yukawa potential and very similar to the combinations of the inversely quadratic Yukawa potential and Yukawa potential on the entire positive line range, $r \in (0, \infty)$. The authors study the solutions of the Dirac equation with the new suggested Yukawa-type potential for any spin-orbit quantum number $j$ interacting with a Coulomb-like tensor interaction. By using the Nikiforov-Uvarov method, Yahya et al. \cite{R12} have recently reported the approximate analytical solutions of the Dirac equation with the shifted Deng-Fan potential including the Yukawa-like tensor interaction under the spin and pseudospin symmetry conditions.
\begin{figure}[!t]
\includegraphics[height=190mm,width=180mm]{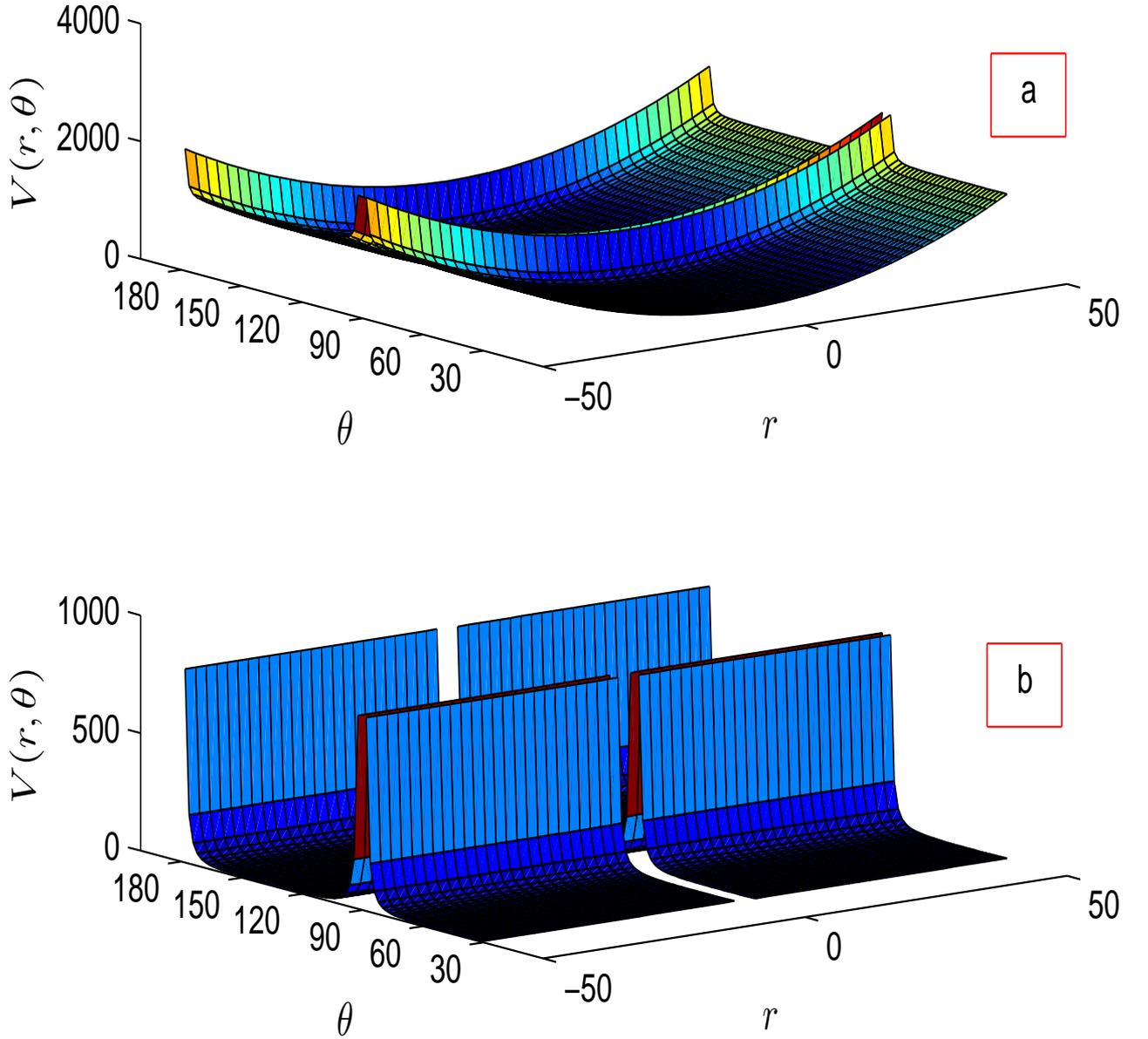}
\caption{{\protect\small (a) The surface plot of double ring shaped oscillator potential as a function of $r$ and $\theta$. We choose $a=b=1$ and $k=1$. (b) The surface plot of double ring shaped Kratzer potential as a function of $r$ and $\theta$. Here we choose $a=b=1$, $r_e=0.4$ and $D_e=12$}}
\label{fig1}
\end{figure}

In the present work, in this section, we attempt to  investigate the Dirac equation under the condition of spin and pseudospin symmetry for the double ring shaped oscillator potential (DRSO) and double ring shaped Kratzer potentials (DRSK) via the asymptotic iteration [78-95,97-102]. These potentials can be described as a three dimensional (3D) harmonic oscillators and Kratzer surrounded by a double ring shaped inverse square potential\cite{YaD07, DuY07}.  In spherical coordinate, the potentials can be written as \cite{YaD07, DuY07, FalThesis}.
\begin{eqnarray}
V(r,\theta)&=&\frac{1}{r^2}\left[\frac{b}{\sin^2\theta}+\frac{a}{\cos^2\theta}\right]+V_{1,2}(r)\ \ \mbox{with}\ \ \left\{\begin{matrix}
V_1(r)&=&-2D_e\left(\frac{r_e}{r}-\frac{1}{2}\frac{r_e^2}{r^2}\right)\ \ \ \mbox{(DRSK),}\\
V_2(r)&=&\frac{kr^2}{2} \ \ \mbox{ (DRSO),}
\end{matrix}\right.
\label{POT1}
\end{eqnarray}
where $k$ denotes the elastic coefficient. $a$ and $b$ are positive real parameters. The surface plot of these two potentials have been presented in figure \ref{fig1}
\section{Dirac equation in view of spin and pseudospin symmetry}
The pseudospin symmetry has been successfully applied to explain different phenomena in nuclear structure since it was introduced in. On the other hand, spin symmetry is relevant for mesons. In spherical coordinates, the Dirac equation for fermionic massive spin$-\frac{1}{2}$ particles interacting with arbitrary scalar potential S(r), the time-component $V(r)$ of a four-vector potential can be expressed as (\cite{OyA12, AyE12, Guo12, ChG12}, and Refs. therein)
\begin{equation}
\left[\vec{\alpha}.\vec{p}+\beta(M+S(r))\right]\psi(\vec{r})=[E-V(\vec{r})]\psi_{n\kappa}(\vec{r}),\ \ \ \ \ \ \psi_{n\kappa}(\vec{r})=\psi_{n\kappa}(r, \theta, \phi),
\label{EQ1}
\end{equation}
where $E$, $\vec{p}$ and $M$ denote the relativistic energy of the system, the momentum operator and mass of the particle respectively. $\alpha$ and $\beta$ are $4\times 4$ Dirac matrices given by
\begin{equation}
\bar{\alpha}=
\left(\begin{array}{lr}     
0&\vec{\sigma }\\      
\vec{\sigma }&0
\end{array}\right)\ ,\ \ \ \beta=
\left(\begin{array}{lr}     
I&0\\      
0&-I
\end{array}\right),\ \ \ {\sigma_1}=
\left(\begin{array}{lr}     
0&1\\      
1&0
\end{array}\right),\ \ \ {\sigma_2}=
\left(\begin{array}{lr}     
0&-1\\      
i&0
\end{array}\right),\ \ \ {\sigma_3}=
\left(\begin{array}{lr}     
1&0\\      
0&-1
\end{array}\right),
\label{EQ2}
\end{equation}
where $I$ is the $2\times2$ unitary matrix and $\vec{\sigma}$  are the three-vector Pauli spin matrices. The eigenvalues of the spin-orbit coupling operator are $\kappa=\left(j+\frac{1}{2}\right)>0$ and $\kappa=-\left(j+\frac{1}{2}\right)<0$ for unaligned spin $j=\ell-\frac{1}{2}$ and the aligned spin $j=\ell+\frac{1}{2}$, respectively. The set $(H^2, K, J^2, J_Z)$ can be taken as the complete set of conservative quantities with $\vec{J}$ being the total angular momentum operator and $K=(\vec{\sigma} .\vec{L}+1)$ is the spin-orbit where $\vec{L}$ is the orbital angular momentum of the spherical nucleons that commutes with the Dirac Hamiltonian\index{Dirac Hamiltonian}. Thus, the spinor wave functions can be classified according to their angular momentum $j$, the spin-orbit quantum number $\kappa$ and the radial quantum number $n$. Hence, they can be written as follows:
\begin{equation}
\psi_{n\kappa}(\vec{r})=
\left(\begin{array}{lr}     
f_{n\kappa}(\vec{r})\\      
g_{n\kappa}(\vec{r})
\end{array}\right),
\label{EQ3}
\end{equation}
where $f_{n\kappa}({\vec{r}})$ and $g_{n\kappa}(\vec{r})$ are the radial wave functions of the upper- and lower-spinor components respectively. Substitution of equation (\ref{EQ3}) into equation (\ref{EQ1}) yields the following coupled differential equations:
\begin{subequations}
\begin{equation}
g_{n\kappa}(\vec{r})=\frac{\sigma.\vec{p}}{\left(E_{n\kappa}+M-\Delta(\vec{r})\right)}f_{n\kappa}(\vec{r}),
\label{EQ4A}
\end{equation}
\begin{equation}
f_{n\kappa}(\vec{r})=\frac{\sigma.\vec{p}}{\left(E_{n\kappa}-M-\sum(\vec{r})\right)}g_{n\kappa}(\vec{r}),
\label{EQ4B}
\end{equation}
\end{subequations}
where $\Delta(\vec{r})=V(\vec{r})-S(\vec{r})$ and $\sum(\vec{r})=V(\vec{r})+S(\vec{r})$ are the difference and sum potentials respectively.  One can eliminate $g_{n\kappa}(\vec{r})$ in equation (\ref{EQ4A}), with the aid of equation (\ref{EQ4B}), to obtain a second-order differential for the upper-spinor component as follows:
\begin{equation}
\left[\frac{(\sigma.\vec{p})(\sigma.\vec{p})}{\left(E_{n\kappa}+M-\Delta(\vec{r})\right)}-\left(E_{n\kappa}-M-\sum(\vec{r})\right)\right]f_{n\kappa}(\vec{r})=0.
\label{EQ5}
\end{equation}
On the other hand, one can eliminate $f_{n\kappa}(\vec{r})$ in equation (\ref{EQ4B}), to obtain a second-order differential equation for the lower-spinor component as follows :
\begin{equation}
\left[\frac{(\sigma.\vec{p})(\sigma.\vec{p})}{\left(E_{n\kappa}-M-\sum(\vec{r})\right)}-\left(E_{n\kappa}+M-\Delta(\vec{r})\right)\right]g_{n\kappa}(\vec{r})=0.
\label{EQ7}
\end{equation}

\subsection{Dirac equation for the radial plus $\Theta$-dependent potentials (Pseudospin symmetry limit)}
Now, we consider radial plus angular-dependent potential $\Delta(\vec{r}) = V_{1,2}(r)+\frac{V(\theta)}{r^2}$. Under the condition of the pseudospin symmetry limit $\frac{d\sum(\vec{r})}{dr} = 0$ or $\sum(\vec{r})=C_{ps}$, equation (\ref{EQ7}) becomes
\begin{equation}
\left[-\vec{\nabla}^2-\left(M+E_{n\kappa}-V(r)-\frac{V(\theta)}{r^2}\right)\left(E_{n\kappa}-M-C_{ps}\right)\right]g_{n\kappa}(\vec{r})=0,
\label{RL1}
\end{equation}
where we have utilized the following relation \cite{Ayd09}
\begin{equation}
\sigma.\vec{p}=\frac{\sigma.\hat{r}}{r}(r\hat{r}.\vec{p}+i\sigma.{}\bf L).
\end{equation}
Now, in order to apply the technique of variable separable, we define \cite{FalThesis}
\begin{equation}
g_{n\kappa}(\vec{r})=\frac{G(r)H(\theta)\Phi(\phi)}{r\sin^{\frac{1}{2}}(\theta)},
\label{RL2}
\end{equation}
and then substitute it into equation (\ref{RL1}) to obtain the following second-order differential equations as follows:
\begin{subequations}
\begin{eqnarray}
\frac{d^2G(r)}{dr^2}+\left[\tilde{\beta^2}-\frac{\tilde{\ell}(\tilde{\ell}+1)}{r^2}-\tilde{\gamma} V_{1,2}(r)\right]G(r)&=&0,
\label{RL3A}\\
\frac{d^2H(\theta)}{d\theta^2}+\left[\left(\tilde{\ell}+\frac{1}{2}\right)^2+\frac{\frac{1}{4}-\tilde{m^2}}{\sin^2\theta}-\tilde{\gamma} V(\theta)\right]H(\theta)&=&0,
\label{RL3B}\\
\frac{d^2\Phi(\phi)}{d\phi^2}+\widetilde{m}^2\Phi(\phi)&=&0,
\label{RL3C}
\end{eqnarray}
\end{subequations}
where we have introduced the following notations for mathematical simplicity
\begin{equation}
\tilde{\gamma}=E_{n\kappa}-M-C_{ps},\ \ \ \ \ \tilde{\beta}^2=(E_{n\kappa}+M)(E_{n\kappa}-M-C_{ps}),
\label{RL4}
\end{equation}
where $E_{n\kappa}\neq M$, i.e., only real negative energy states exist when $C_{ps}=0$(exact pseudospin symmetry). From the above equations, the energy eigenvalues depend on the quantum number $n$ and $\kappa$ (i.e., $E_{n\kappa}=E(n,\kappa(\kappa-1))$), and also the pseudo-orbital angular quantum number $\tilde{\ell}$ according to $\kappa(\kappa-1)=\tilde{\ell}(\tilde{\ell}+1)$ which implies that $j=\tilde{\ell}\pm 1/2$ are degenerate for $\tilde{\ell}\neq 0$. The quantum condition is obtained from the finiteness of the solution at infinity and at the origin point, i.e. , $F_{n\kappa}(0)=G_{n\kappa}(0)=0$ and $F_{n\kappa}(\infty)=G_{n\kappa}(\infty)=0$.

\subsection{Dirac equation for the radial plus $\Theta$-dependent potentials (Spin symmetry limit)}
Again, we consider radial plus angular-dependent potential $\sum(\vec{r}) = V_{1,2}(r)+\frac{V(\theta)}{r^2}$. In this case, under the condition of the spin limit $\frac{d\Delta(\vec{r})}{dr} = 0$ or $\Delta(\vec{r})=C_s$, equation (\ref{EQ5}) becomes:
\begin{equation}
\left[-\vec{\nabla}^2-\left(M-E_{n\kappa}+V_{1,2}(r)+\frac{V(\theta)}{r^2}\right)\left(C_{s}-M-E_{n\kappa}\right)\right]f_{n\kappa}(\vec{r})=0,
\label{EQ9}
\end{equation}
Then we apply method of variable separable by defining \cite{FalThesis}
\begin{equation}
f_{n\kappa}(\vec{r})=\frac{F(r)H(\theta)\Phi(\phi)}{r\sin^{\frac{1}{2}}(\theta)}.
\label{EQ10}
\end{equation}
Then, the following second-order differential equations are obtained as follows:
\begin{subequations}
\begin{eqnarray}
\frac{d^2F(r)}{dr^2}+\left[\beta^2-\frac{\ell(\ell+1)}{r^2}-\gamma V_{1,2}(r)\right]F(r)&=&0,
\label{EQ11A}\\
\frac{d^2H(\theta)}{d\theta^2}+\left[\left(\ell+\frac{1}{2}\right)^2+\frac{\left(\frac{1}{4}-m^2\right)}{\sin^2\theta}-\gamma V(\theta)\right]H(\theta)&=&0,
\label{EQ11B}\\
\frac{d^2\Phi(\phi)}{d\phi^2}+m^2\Phi(\phi)&=&0,
\label{EQ11C}
\end{eqnarray}
\end{subequations}
where we have introduced the following notations for mathematical simplicity: 
\begin{equation}
\gamma=E_{n\kappa}+M-C_s,\ \ \ \ \ \beta^2=(E_{n\kappa}-M)(C_s-E_{n\kappa}-M),
\label{EQ12}
\end{equation}
where $E_{n\kappa}\neq-M$, i.e., only real positive energy states exist when $C_s=0$(exact spin symmetry). $\kappa(\kappa+1)=\ell(\ell+1)$, $\kappa =\ell$ for $\kappa<0$ and $\kappa=-(\ell+1)$ for $\kappa>0$. The spin symmetry energy depend on $n$ and $\kappa$, i.e., $E_{n\kappa}=E(n,\kappa(\kappa+1))$. For $\ell\neq0$, the states with $j=\ell\pm1/2$ are degenerate. $m$ and $\ell$ are separation constants. The relation between $\kappa$ and $\ell$ is given as
\begin{eqnarray}
\kappa=
\left\{\begin{array}{llcrr}     
-(\ell+1) =\left(j+\frac{1}{2}\right), &(s_{1/2},p_{3/2},etc.)\ \ j=\ell+\frac{1}{2},\ \ \mbox{aligned spin}\ \ (\kappa<0)\\   
+\ell=+\left(j+\frac{1}{2}\right),  &(p_{1/2},d_{3/2},etc.),\ \ j=\ell-\frac{1}{2},\ \ \mbox{unaligned spin}\ \ (\kappa>0)
\end{array}\right.
\end{eqnarray}
\section{Pseudospin symmetry solutions of the double ring-shaped oscillator and Kratzer potential}
Firstly, we investigate the pseudospin symmetry by considering equations (\ref{EQ11A}), (\ref{EQ11B}) and (\ref{EQ11C}). With potential (\ref{POT1}), the equations can be re-written as
\begin{subequations}
\begin{eqnarray}
\frac{d^2G(r)}{dr^2}+\left[\tilde{\beta}^2-\frac{\tilde{\ell}(\tilde{\ell}+1)}{r^2}-\tilde{\gamma}V_{1,2}(r)\right]G(r)&=&0,\label{EQ21A}\\
\frac{d^2H(\theta)}{d\theta^2}+\left[\left(\tilde{\ell}+\frac{1}{2}\right)^2+\frac{\left(\frac{1}{4}-\widetilde{m}^2\right)}{\sin^2\theta}-\tilde{\gamma} \left[\frac{b}{\sin^2\theta}+\frac{a}{\cos^2\theta}\right]\right]H(\theta)&=&0,
\label{EQ21B}\\
\frac{d^2\Phi(\phi)}{d\phi^2}+\widetilde{m}^2\Phi(\phi)&=&0,
\label{EQ21C}
\end{eqnarray}
\end{subequations}
where $\widetilde{m}^2$ and $\tilde{\ell}(\tilde{\ell}+1)$ are separation constants. The components of the wave function required to satisfy the boundary conditions of $G(0)=0$ and $G(\infty)=0$ in equation (\ref{EQ21A}), $H(\theta)$ and $H(\pi)$ are infinite in equation (\ref{EQ21B}) and 
$\Phi(\phi)=\Phi(\phi+2\pi)$ in equation (\ref{EQ21C}). On considering a case where $V(\phi)=0$, the normalized solution of equation (\ref{EQ21C}) can be written as
\begin{equation}
\Phi_{\widetilde{m}}(\phi)=\frac{exp(i\widetilde{m}\phi)}{\sqrt{2\pi}}\ \ \ \ \ \ \ (\widetilde{m}=0, \pm 1, \pm 2, \pm 3, .....),
\label{EQ22}
\end{equation}
which provides the boundary condition. Equations (\ref{EQ21A}) and (\ref{EQ21B}) are radial and polar-angle equations which we shall solved in this section later. 
\subsection{CASE 1, when $V_1(r)=-2D_e\left(\frac{r_e}{r}-\frac{1}{2}\frac{r_e^2}{r^2}\right)$: Pseudospin symmetry solutions of the double ring-shaped Kratzer potential}
Let us first consider the radial part of the problem by solving equation (\ref{EQ21A}) analytically. Thus we can have
\begin{equation}
\frac{d^2G(r)}{dr^2}+\left[\widetilde{\beta}^2-\frac{\widetilde{\ell}(\widetilde{\ell}+1)+\widetilde{\gamma} D_er_e^2}{r^2}+\frac{2\widetilde{\gamma} D_er_e}{r}\right]G(r)=0.
\label{EQ23}
\end{equation}
Equation (\ref{EQ23}) can be solved by using asymptotic iteration method by considering the boundary conditions first and then consider the asymptotic behavior of the radial wave function as $r\rightarrow 0$ and $r\rightarrow \infty$. Thus, the physically acceptable solution to the $G(r)$ can be expressed as
\begin{equation}
G(r)= r^{\widetilde{\zeta}} e^{-\sqrt{-\widetilde{\beta}^2}r}R(r), \ \ \ \mbox{with} \ \ {\widetilde{\zeta}}=\frac{1}{2}+\sqrt{\left(\widetilde{\ell}+\frac{1}{2}\right)^2+\widetilde{\gamma}D_er_e^2}.
\label{EQ24}
\end{equation}
On putting this wave function into equation (\ref{EQ23}), we obtain the following second-order homogeneous linear differential equation which is suitable to solve with the asymptotic iteration method
\begin{equation}
R''(r)+2\left(\frac{{\widetilde{\zeta}}}{r}-\sqrt{-\widetilde{\beta}^2}\right)R'(r)+2\left(\frac{\widetilde{\gamma}D_er_e-{\widetilde{\zeta}}\sqrt{-\widetilde{\beta}^2}}{r}\right)R(r)=0.
\label{EQ25}
\end{equation}
By comparing equation (\ref{EQ25}) with equation (\ref{A4}) we can easily find $\lambda_0(r)$ and $s_0(r)$ values. Thus, with the help of recurrence relation (\ref{A4}) and termination condition (\ref{A5}), it is straightforward to obtain the following series:
\begin{eqnarray}
&&\delta_0(r)=
\left|
\begin{matrix}     
\lambda_0(r)&s_0(r) \\      
  \lambda_{1}(r)&s_{1}(r)
  \end{matrix}
  \right|=0\ \ \ \ \ \ \ \Rightarrow\ \ \ \ \ \  -\sqrt{-\widetilde{\beta}^2}=\frac{\widetilde{\gamma}D_er_e}{{\widetilde{\zeta}}+0}\nonumber\\
&&\delta_1(r)=
\left|
\begin{matrix}     
\lambda_1(r)&s_1(r) \\      
  \lambda_{2}(r)&s_{2}(r)
  \end{matrix}
  \right|=0\ \ \ \ \ \ \ \Rightarrow\ \ \ \ \ \  -\sqrt{-\widetilde{\beta}^2}=\frac{\widetilde{\gamma}D_er_e}{{\widetilde{\zeta}}+1}\nonumber\\
	&&\delta_2(r)=
\left|
\begin{matrix}     
\lambda_2(r)&s_2(r) \\      
  \lambda_{3}(r)&s_{3}(r)
  \end{matrix}
  \right|=0\ \ \ \ \ \ \ \Rightarrow\ \ \ \ \ \  -\sqrt{-\widetilde{\beta}^2}=\frac{\widetilde{\gamma}D_er_e}{{\widetilde{\zeta}}+2}\\
 &&\delta_3(r)=
\left|
\begin{matrix}     
\lambda_3(r)&s_3(r) \\      
  \lambda_{4}(r)&s_{4}(r)
  \end{matrix}
  \right|=0\ \ \ \ \ \ \ \Rightarrow\ \ \ \ \ \  -\sqrt{-\widetilde{\beta}^2}=\frac{\widetilde{\gamma}D_er_e}{{\widetilde{\zeta}}+3}\nonumber\\
	&&.... etc.\nonumber
	\label{EQ26}
\end{eqnarray}
By finding the nth term of the above series, and substitute for the values of parameter $\widetilde{\beta}$, $\widetilde{\gamma}$, ${\widetilde{\zeta}}$, we obtain an explicit expression for the relativistic energy equation as:
\begin{equation}
\frac{E_{n\kappa}+M}{M-E_{n\kappa}+C_{ps}}=\frac{D_e^2r_e^2}{\left(n+\frac{1}{2}+\sqrt{\left(\widetilde{\ell}+\frac{1}{2}\right)^2+\left(E_{n\kappa}-M-C_{ps}\right)D_er_e^2}\right)^2}.
\label{EQ27}
\end{equation}
By following the procedure described by Appendix B, the radial wave function can be obtained as
\begin{equation}
G(r)=(-1)^n\frac{\Gamma\left(2\sqrt{-\widetilde{\beta}^2}+n\right)}{\Gamma\left(2\sqrt{-\widetilde{\beta}^2}\right)}N_ne^{-\sqrt{-\widetilde{\beta}^2}r}\ _1F_1\left(-n, 2{\widetilde{\zeta}}, 2\sqrt{-\widetilde{\beta}^2}r\right),
\label{EQ28}
\end{equation}
where we have used the following limit expression of the hypergeometric function \cite{{Fal124}}:
\begin{equation}
\lim_{b\rightarrow 0}\ _2F_1\left(-n; \frac{1}{b}+a; c; xb\right)=_1F_1(-n; c; x),\ \ y_n(x)=(-1)^nC_2(N+2)^n\ _1F_1\left(-n, \sigma; \frac{2a}{N+2}x^{N+2}\right),
\label{EQ29}
\end{equation}
and $N_n$ is the normalization constant. The solution of the angular part can be obtained by introducing a new transformation of the form $x=\sin^2\theta$ in equation (\ref{EQ21B}), so that we have
\begin{equation}
\frac{d^2H(x)}{dx^2}+\frac{\frac{1}{2}-x}{x(1-x)}\frac{dH(x)}{dx}+\frac{1}{4x(1-x)}\left[\left(\widetilde{\ell}+\frac{1}{2}\right)^2+\frac{\frac{1}{4}-\widetilde{m}^2-\widetilde{\gamma}b}{x}-\frac{\widetilde{\gamma}a}{1-x}\right]H(x)=0.
\label{EQ30}
\end{equation}
According to the Frebenius theorem, the singularity points of the above differential equation (\ref{EQ30}) play an important role in the form of the wave functions. The singular points here are at $x=0$ $(\theta=0)$ and at $x=1$ $(\theta=\frac{\pi}{2})$. As a result, we take the wave function of the form 
\begin{equation}
H(x)=x^{\widetilde{\eta}}(1-x)^{\widetilde{p}}\bar{H}(x), \ \ \mbox{with}\ \ \ \widetilde{\eta}=\frac{1}{4}\left[1+2\sqrt{\widetilde{m}^2+\widetilde{\gamma}b}\right],\ \ \ \widetilde{p}={\frac{1}{4}\left[1+2\sqrt{\frac{1}{4}+\widetilde{\gamma}a}\right]}.
\label{EQ31}
\end{equation}
With equation (\ref{EQ31}), equation (\ref{EQ30}) can be transformed into a more convenient equation that is suitable to apply the asymptotic iteration method
\begin{equation}
\bar{H}''(x)+\left[\frac{\left(2\widetilde{\eta}+\frac{1}{2}\right)-x(\widetilde{\eta}+\widetilde{p}+1)}{x(1-x)}\right]\bar{H}'(x)-\left[\frac{(\widetilde{\eta}+\widetilde{p})^2-\frac{1}{4}\left(\widetilde{\ell}+\frac{1}{2}\right)^2}{x(1-x)}\right]\bar{H}(x)=0.
\label{EQ32}
\end{equation}
By comparing equation (\ref{EQ32}) with equation (\ref{A4}), we can obtain $\lambda_0$, $s_0$ and consequently, $\lambda_1$, $s_1$, $\lambda_2$, $s_2$ ...... $\lambda_k$, $s_k$ can be easily obtained. Thus using the termination condition given by appendix (\ref{A5}), we obtain the following series
\begin{eqnarray}
&&s_0\lambda_1-s_1\lambda_0=0\ \ \ \ \Leftrightarrow\ \ \ \ \ \widetilde{\eta}+\widetilde{p}=\ \ \ 0+\frac{1}{2}\left({\widetilde{\ell}+\frac{1}{2}}\right)\nonumber\\
&&s_1\lambda_2-s_2\lambda_1=0\ \ \ \ \Leftrightarrow\ \ \ \ \ \widetilde{\eta}+\widetilde{p}=-1+\frac{1}{2}\left({\widetilde{\ell}+\frac{1}{2}}\right)\\
&&s_2\lambda_3-s_3\lambda_2=0\ \ \ \ \Leftrightarrow\ \ \ \ \ \widetilde{\eta}+\widetilde{p}=-2+\frac{1}{2}\left({\widetilde{\ell}+\frac{1}{2}}\right)\nonumber\\
&&s_3\lambda_4-s_4\lambda_3=0\ \ \ \ \Leftrightarrow\ \ \ \ \ \widetilde{\eta}+\widetilde{p}=-3+\frac{1}{2}\left({\widetilde{\ell}+\frac{1}{2}}\right)\nonumber\\
&&.... etc.\nonumber
\label{EQ33}
\end{eqnarray}
When the above expressions are generalized then substitute for $\widetilde{\eta}$, $\widetilde{p}$, $\widetilde{\beta}$ and $\widetilde{\gamma}$, the eigenvalues equation turns to
\begin{equation}
\widetilde{\ell}+\frac{1}{2}=\sqrt{a\left(E_{n\kappa}-M-C_{ps}\right)+\frac{1}{4}}+\sqrt{b\left(E_{n\kappa}-M-C_{ps}\right)+\widetilde{m}^2}+2n'+1,
\label{EQ34}
\end{equation}
where $n'=0, 1, 2, ......$. The wave function can also be found as
\begin{equation}
H(\theta)=(-1)^nN_{n'\widetilde{m}}\frac{\Gamma(2\widetilde{\eta}+1/2+n)}{\Gamma(2\widetilde{\eta}+1/2)}\sin^{2\widetilde{\eta}}\theta\cos^{2\widetilde{p}}\theta\ _2F_1\left(-n', n'+2(\widetilde{\eta}+\widetilde{p}); 2\widetilde{\eta}+\frac{1}{2}, \sin^2\theta\right),
\label{EQ35}
\end{equation}
where $N_{n'\widetilde{m}}$ is the normalization constant. By replacing $\tilde{\ell}+1/2$ in equation (\ref{EQ27}) by equation (\ref{EQ34}), we can find the relativistic rovibrational energy spectrum for a diatomic molecule system in the presence of a double ring-shaped Kratzer potential as
\begin{eqnarray}
\frac{E_{n\widetilde{m}n'}^{(DRSK)}+M}{M-E_{n\widetilde{m}n'}+C_{ps}}=\frac{D_e^2r_e^2}{\left(n+\frac{1}{2}+\sqrt{\left(\Omega(\widetilde{\ell})+2n'+1\right)^2+\left(E_{n\widetilde{m}n'}^{(DRSK)}-M-C_{ps}\right)D_er_e^2}\right)^2},\ \ \mbox{with} \nonumber\\
\Omega(\widetilde{\ell})=\sqrt{a\left(E_{n\widetilde{m}n'}^{(DRSK)}-M-C_{ps}\right)+\frac{1}{4}}+\sqrt{b\left(E_{n\widetilde{m}n'}^{(DRSK)}-M-C_{ps}\right)+\widetilde{m}^2}
\label{EQ36}
\end{eqnarray}
and the complete wave function as
\begin{eqnarray}
g_{n\widetilde{m}n'}^{(DRSK)}(\vec{r})&=&(-1)^{2n}\frac{N_{n\widetilde{m}n'}}{r\sin^{\frac{1}{2}}\theta}\frac{\Gamma(2\widetilde{\eta}+1/2+n)}{\Gamma(2\widetilde{\eta}+1/2)}\frac{\Gamma\left(2\sqrt{-\widetilde{\beta}^2}+n\right)}{\Gamma\left(2\sqrt{-\widetilde{\beta}^2}\right)}\frac{exp(im\phi)}{\sqrt{2\pi}}r^{\widetilde{\zeta}} e^{-\sqrt{-\widetilde{\beta}^2}r}\sin^{2\widetilde{\eta}}\theta\cos^{2\widetilde{p}}\theta \nonumber\\
&\times&\ _1F_1\left(-n, 2\widetilde{\zeta}, 2\sqrt{-\widetilde{\beta}^2}r\right)\ _2F_1\left(-n', n'+2(\widetilde{\eta}+\widetilde{p}); 2\eta+\frac{1}{2}, \sin^2\theta\right),
\label{EQ37}
\end{eqnarray}
where $N_{nmn'}$ is the normalization constant.
\subsection{CASE 2, when $V_2(r)=\frac{kr^2}{2}$: Pseudospin symmetry solutions of the double ring-shaped oscillator potential}
We begin by considering the radial part of the problem. we can now write equation (\ref{EQ11A}) as
\begin{equation}
\frac{d^2G(r)}{dr^2}+\left[\widetilde{\beta}^2-\frac{\widetilde{\gamma}kr^2}{2}-\frac{\widetilde{\ell}(\widetilde{\ell}+1)}{r^2}\right]G(r)=0.
\label{EQ38}
\end{equation}
In order to make equation (\ref{EQ38}) suitable to apply AIM, we introduce a new transformation of the form $s=r^2$ with which equation (\ref{EQ38}) reduces to
\begin{equation}
4sG''(s)+2G'(s)+\left[\widetilde{\beta}^2-\frac{\widetilde{\gamma}ks}{2}-\frac{\widetilde{\ell}(\widetilde{\ell}+1)}{s}\right]G(s)=0.
\label{EQ39}
\end{equation}
The above equation (\ref{EQ39}) has irregular singularity as $s\rightarrow\infty$, where its normalized solutions in bound states behave like $e^{-\sqrt{-\frac{\widetilde{\gamma}k}{8}}}$. Furthermore, it has a singularity as $r\rightarrow 0$, where $G(s)\rightarrow s^{\ell+1}$. Therefore the reasonable wave function we proposed is as follows:
\begin{equation}
G(s)=s^{\frac{1}{2}(\widetilde{\ell}+1)}e^{-\sqrt{-\frac{\widetilde{\gamma}k}{8}}}R(s)
\label{EQ40}
\end{equation}
The solution of equation (\ref{EQ39}) can therefore be easily found by using asymptotic iteration method as
\begin{equation}
\left(M+E_{n\kappa}\right)\sqrt{\left(E_{n\kappa}-M-C_{ps}\right)}=\sqrt{-2k}\left(\widetilde{\ell}+\frac{3}{2}+2n\right),
\label{EQ41}
\end{equation}
and the wave function as
\begin{equation}
G(r)=(-1)^n\frac{\Gamma\left(\widetilde{\ell}+n+\frac{3}{2}\right)}{\Gamma\left(\widetilde{\ell}+\frac{3}{2}\right)}N_nr^{(\widetilde{\ell}+1)}e^{-\sqrt{-\frac{\widetilde{\gamma}k}{8}}r^2}\ _1F_1\left(-n, \widetilde{\ell}+\frac{3}{2}, r^2\sqrt{-\frac{\widetilde{\gamma}k}{2}} \right),
\label{EQ42}
\end{equation}
\begin{table}[!t]
{\scriptsize
\caption{ \small The pseudospin bound state energy eigenvalues of the Dirac equation in units of $fm^{-1}$ with the Double ring shaped Kratzer potential } \vspace*{10pt}{
\begin{tabular}{ccccccccc}\hline\hline
{}&{}&{}&{}&{}&{}&{}&{}&{}\\[-1.0ex]
$n$	&$\widetilde{n}$	&$\widetilde{m}$&$a=b=1$&$a=0, b=1$&$a=1, b=0$&$a=b=0$	\\[1ex]\hline\hline
0	&0	&0	&-1.470943351	&-0.016009740	&-0.235179025 &-0.361711704,  1.666666667\\[1ex]
1	&0	&0	&-2.431421872	&-1.567942947	&-1.102586154	&-0.653238514,  0.772422545\\[1ex]
1	&0	&1	&-2.469116554	&-1.625811990	&-2.067327828	&-1.185623781,  0.053405584\\[1ex]
1	&1	&0	&-3.699882631	&-3.533344802	&-3.063074782	&-1.722141025, -1.116384727\\[1ex]
1	&1	&1	&-3.757254870	&-3.635701770	&-4.036449245	&-2.220469470, -2.726621983\\[1ex]
2	&0	&0	&-3.209457737	&-2.827950086	&-2.449594892	&-0.898099602, -1.219724870\\[1ex]
2	&0	&1	&-3.255821962	&-2.889117859	&-3.244411758	&-1.876423389, -1.576083657\\[1ex]
2	&1	&0	&-4.245687715	&-4.446330720	&-4.065979868	&-2.989578899, -2.131084520\\[1ex]
2	&1	&1	&-4.304647881	&-4.548045830	&-4.857898865	&-4.557401104, -2.599289725\\[1ex]
2	&2	&0	&-5.017793076	&-5.807805499	&-5.551416345	&-2.985623436, -6.568783099\\[1ex]
2	&2	&1	&-5.083773897	&-5.939848799	&-6.076808748	&-3.299901860, -9.015137661\\[1ex]
3	&0	&0	&-3.840151551	&-3.825089370	&-3.523270148	&-2.821605268\\[1ex]
3	&0	&1	&-3.891267802	&-3.888870745	&-4.165722690	&-1.934978525, -3.424312038\\[1ex]
3	&1	&0	&-4.672598850	&-5.122648167	&-4.826237022	&-4.475508691, -2.500000000\\[1ex]
3	&1	&1	&-4.731057731	&-5.222159573	&-5.446999890	&-2.929166516, -5.993063122\\[1ex]
\hline\hline
\end{tabular}\label{Tab2}}
\vspace*{-1pt}}
\end{table}
\begin{table}[!t]
{\scriptsize
\caption{ \small The pseudospin bound state energy eigenvalues of the Dirac equation in units of $fm^{-1}$ with the Double ring shaped oscillator potential } \vspace*{10pt}{
\begin{tabular}{ccccccccc}\hline\hline
{}&{}&{}&{}&{}&{}&{}&{}&{}\\[-1.0ex]
$n$	&$\widetilde{n}$	&$\widetilde{m}$&$a=b=1$&$a=0, b=1$&$a=1, b=0$&$a=b=0$	\\[1ex]\hline\hline
0	&0	&0	&-3.788148054	&-3.237560428	&-3.594587451	&-0.6652434115		\\[1ex]
1	&0	&0	&-2.699991845	&-2.036528960	&-2.212632857	&-1.3261285500		\\[1ex]
1	&0	&1	&-2.552209168	&-1.841659094	&-1.799313908	&-1.1063485860	  \\[1ex]
1	&1	&0	&-1.914303457	&-1.368854413	&-1.477685069	&-0.8957672743		\\[1ex]
1	&1	&1	&-1.793428823	&-1.236923669	&-1.202340221	&-0.6930492526		\\[1ex]
2	&0	&0	&-1.914303457	&-1.368854413	&-1.477685069	&-0.8957672743		\\[1ex]
2	&0	&1	&-1.793428823	&-1.236923669	&-1.202340221	&-0.6930492526		\\[1ex]
2	&1	&0	&-1.319244086	&-0.863791517	&-0.954556403	&-0.4971622035		\\[1ex]
2	&1	&1	&-1.221377650	&-0.761977326	&-0.725329392	&-0.3072877327		\\[1ex]
2	&2	&0	&-0.822355777	&-0.427394429	&-0.509690516	&-0.1227634742		\\[1ex]
2	&2	&1	&-0.740004577	&-0.343059305	&-0.304599059	& 0.0569560111		\\[1ex]
3	&0	&0	&-1.319244086	&-0.863791517	&-0.954556403	&-0.4971622035		\\[1ex]
3	&0	&1	&-1.221377650	&-0.761977326	&-0.725329392	&-0.3072877327		\\[1ex]
3	&1	&0	&-0.822355777	&-0.427394429	&-0.509690516	&-0.1227634742		\\[1ex]
3	&1	&1	&-0.740004577	&-0.343059305	&-0.304599059	& 0.0569560111		\\[1ex]
\hline\hline
\end{tabular}\label{Tab2}}
\vspace*{-1pt}}
\end{table}
where $N_n$ is the normalization constant. By inserting $\widetilde{\ell}+\frac{1}{2}$ given by equation (\ref{EQ34}) into equation (\ref{EQ41}), the relativistic energy spectrum for a bound electron in the presence of a ring-shaped potential can be found as
\begin{equation}
\frac{\left(M+E_{n\widetilde{m}n'}^{(DRSO)}\right)\sqrt{\left(E_{n\widetilde{m}n'}^{(DRSO)}-M-C_{ps}\right)}}{\left(\sqrt{a\left(E_{n\widetilde{m}n'}^{(DRSO)}-M-C_{ps}\right)+\frac{1}{4}}+\sqrt{b\left(E_{n\widetilde{m}n'}^{(DRSO)}-M-C_{ps}\right)+\widetilde{m}^2}+2n'+2+2n\right)}=\sqrt{-2k},
\label{EQ43}
\end{equation}
and the complete wave functions as
\begin{eqnarray}
g_{n\widetilde{m}n'}^{(DRSO)}(\vec{r})&=&(-1)^{2n}\frac{C_{n\widetilde{m}n'}}{r\sin^{\frac{1}{2}}\theta}\frac{\Gamma(2\widetilde{\eta}+1/2+n)}{\Gamma(2\widetilde{\eta}+1/2)}\frac{\Gamma\left(\widetilde{\ell}+n+\frac{3}{2}\right)}{\Gamma\left(\widetilde{\ell}+\frac{3}{2}\right)} \frac{exp(i\widetilde{m}\phi)}{\sqrt{2\pi}}r^{(\widetilde{\ell}+1)}e^{-\sqrt{-\frac{\widetilde{\gamma}k}{8}}r^2}\sin^{2\widetilde{\eta}}\theta\cos^{2\widetilde{p}}\theta\ \nonumber\\
&\times&_1F_1\left(-n, \widetilde{\ell}+\frac{3}{2}, r^2\sqrt{-\frac{\widetilde{\gamma}k}{2}} \right)\ _2F_1\left(-n, n+2(\widetilde{\eta}+\widetilde{p}); 2\widetilde{\eta}+\frac{1}{2}, \sin^2\theta\right),
\label{EQ44}
\end{eqnarray}
where $C_{n\widetilde{m}n'}$ is the normalization factor for the complete wavefunction.
\section{Spin symmetry solutions of the double ring-shaped oscillator and Kratzer potential}
Now let investigate the spin symmetry solutions of the problems by considering equations (\ref{EQ11A}), (\ref{EQ11B}) (\ref{EQ11C}) and potential equation (\ref{POT1}) to have
\begin{subequations}
\begin{eqnarray}
\frac{d^2F(r)}{dr^2}+\left[{\beta}^2-\frac{{\ell}({\ell}+1)}{r^2}-{\gamma}V_{1,2}(r)\right]F(r)&=&0,\\
\label{EQ45A}
\frac{d^2H(\theta)}{d\theta^2}+\left[\left({\ell}+\frac{1}{2}\right)^2+\frac{\left(\frac{1}{4}-{m}^2\right)}{\sin^2\theta}-{\gamma} \left[\frac{b}{\sin^2\theta}+\frac{a}{\cos^2\theta}\right]\right]H(\theta)&=&0,
\label{EQ45B}\\
\frac{d^2\Phi(\phi)}{d\phi^2}+{m}^2\Phi(\phi)&=&0,
\label{EQ45C}
\end{eqnarray}
\end{subequations}
\begin{table}[!t]
{\scriptsize
\caption{ \small The spin bound state energy eigenvalues of the Dirac equation in units of $fm^{-1}$ with the Double ring shaped Kratzer potential } \vspace*{10pt}{
\begin{tabular}{ccccccccc}\hline\hline
{}&{}&{}&{}&{}&{}&{}&{}&{}\\[-1.0ex]
$n$	&$\widetilde{n}$	&$\widetilde{m}$&$a=b=1$&$a=0, b=1$&$a=1, b=0$&$a=b=0$	\\[1ex]\hline\hline
0	&0	&0	&2.072188142	&12.09217551	&1.116712576	&17.29953766, 0.744179704	\\[1ex]
	    &&	&9.060994522	&1.406939539	&12.74519748	&-0.400946639	\\[1ex]
1	&0	&0	&2.725765193	&2.166121215	&1.947036165	&1.493268566, 13.98804932	\\[1ex]
	    &&	&8.207625097	&10.32429829	&10.81287612	&-0.751271606	\\[1ex]
1	&0	&1	&8.103584648	&10.16638485	&2.447369328	&13.07639776, 1.955144908	\\[1ex]
    	&&	&2.845560703	&2.315560481	&9.779027300	&-1.66666667	\\[1ex]
1	&1	&0	&7.186557630	&3.012069404	&8.868839382	&11.87959406, 2.389717500	\\[1ex]
    	&&	&3.425589261	&8.504093195	&2.860071672	&-3.079630218	\\[1ex]
1	&1	&1	&3.490508001	&8.378950515	&8.115160360	&10.59130823, 2.769682610	\\[1ex]
	    &&	&7.113819334	&3.099115024	&3.193384975	&-5.205436911	\\[1ex]
2	&0	&0	&3.167137607	&2.720904245	&2.561073200	&11.53551968, 2.156404231	\\[1ex]
	    &&	&7.561368032	&9.033462867	&9.389148000	&	\\[1ex]
2	&0	&1	&3.247127506	&2.820166862	&2.922513885	&10.82905932,  2.500000000	\\[1ex]
	    &&	&7.479670321	&8.910883178	&8.634882758	&-2.292912051	\\[1ex]
2	&1	&0	&6.787568957	&7.717297246	&3.230176780	&9.943446173, -4.123924146 	\\[1ex]
	    &&	&3.678964621	&3.346561778	&7.978234688	&2.834669759	\\[1ex]
2	&1	&1	&3.727193112	&3.410443771	&3.484971875	&9.03830004, 3.132086660	\\[1ex]
	    &&	&6.730700771	&7.623294975	&7.436562470	&-6.702239952	\\[1ex]
2	&2	&0	&6.281221374	&6.839147059	&7.003717432	&8.23583808, 3.38716421	\\[1ex]
	    &&	&4.010999391	&3.774204752	&3.694652628	&-10.17752526	\\[1ex]
2	&2	&1	&4.041936969	&3.815408674	&3.867483358	&7.58674172, 3.60267187 	\\[1ex]
	    &&	&6.242991096	&6.776804599	&6.662707035	&-14.61359280	\\[1ex]
3	&0	&0	&3.485915060	&8.106020479	&8.362128945	&9.795375780, 2.678986992	\\[1ex]
	    &&	&7.073115012	&3.131267848	&3.012356170	&	\\[1ex]
3	&0	&1	&7.009267661	&3.201037907	&7.816722460	&9.272753313, 2.932318925 	\\[1ex]
	    &&	&3.543052442	&8.012834667	&3.280500248	&-2.604114296	\\[1ex]
3	&1	&0	&6.482650213	&3.605610390	&7.341939088	&8.640638643, 3.188810550	\\[1ex]
	    &&	&3.874907652	&7.153253654	&3.514963518	&-4.888919080	\\[1ex]
3	&1	&1	&3.911932694	&7.082443862	&6.947883890	&8.01410686,  3.421849740	\\[1ex]
			&&  &6.437892252	&3.653932542	&3.713426640	&-7.848088905	\\[1ex]
\hline\hline
\end{tabular}\label{Tab2}}
\vspace*{-1pt}}
\end{table}
\begin{table}[!t]
{\scriptsize
\caption{ \small The spin bound state energy eigenvalues of the Dirac equation in units of $fm^{-1}$ with the Double ring shaped oscillator potential } \vspace*{10pt}{
\begin{tabular}{ccccccccc}\hline\hline
{}&{}&{}&{}&{}&{}&{}&{}&{}\\[-1.0ex]
$n$	&$\widetilde{n}$	&$\widetilde{m}$&$a=b=1$&$a=0, b=1$&$a=1, b=0$&$a=b=0$	    \\[1ex]\hline\hline
0	&0	&0	&5.355386495	&5.344894530  &5.275966345	&5.212382260, -0.424764518		\\[1ex]
1	&0	&0	&5.722284130	&5.698887675	&5.622916065	&5.545734075, -1.091468151		\\[1ex]
1	&0	&1	&5.777138920	&5.752777705	&5.803291465	&5.726461280, -1.452922562		\\[1ex]
1	&1	&0	&6.086063615	&6.060138650	&5.983656065	&5.908930160, -1.817860321		\\[1ex]
1	&1	&1	&6.136772850	&6.110900805	&6.162525320	&6.090698405, -2.181396810  	\\[1ex]
2	&0	&0	&6.086063615	&6.060138650	&5.983656065	&5.908930160, -1.817860321		\\[1ex]
2	&0	&1	&6.136772850	&6.110900805	&6.162525320	&6.090698405, -2.181396810  	\\[1ex]
2	&1	&0	&6.439303895	&6.414206035	&6.339134305	&6.270509265, -2.541018528		\\[1ex]
2	&1	&1	&6.486245415	&6.461558695	&6.513111705	&6.447739155, -2.895478312		\\[1ex]
2	&2	&0	&6.780625095	&6.757438625	&6.684305200	&6.622108225, -3.244216452		\\[1ex]
2	&2	&1	&6.824308155	&6.801654945	&6.852685530	&6.793525590, -3.587051182		\\[1ex]
3	&0	&0	&6.439303895	&6.414206035	&6.339134305	&6.270509265, -2.541018528		\\[1ex]
3	&0	&1	&6.486245415	&6.461558695	&6.513111705	&6.447739155, -2.895478312		\\[1ex]
3	&1	&0	&6.780625095	&6.757438625	&6.684305200	&6.622108225, -3.244216452		\\[1ex]
3	&1	&1	&6.824308155	&6.801654945	&6.852685530	&6.793525590, -3.587051182		\\[1ex]
\hline\hline
\end{tabular}\label{Tab2}}
\vspace*{-1pt}}
\end{table}
where ${m}^2$ and ${\ell}({\ell}+1)$ are separation constants. The solution of equation (\ref{EQ45C}) is also found as (\ref{EQ22}) with transformation $m=\widetilde{m}$
\subsection{CASE 1, when $V_1(r)=-2D_e\left(\frac{r_e}{r}-\frac{1}{2}\frac{r_e^2}{r^2}\right)$: Spin symmetry solutions of the double ring-shaped Kratzer potential}
By considering the radial and angular part of the problem, we can have the following equations
\begin{eqnarray}
\frac{d^2F(r)}{dr^2}+\left[{\beta}^2-\frac{{\ell}({\ell}+1)+{\gamma} D_er_e^2}{r^2}+\frac{2{\gamma} D_er_e}{r}\right]F(r)=0,\nonumber\\
\frac{d^2H(\theta)}{d\theta^2}+\left[\left({\ell}+\frac{1}{2}\right)^2+\frac{\left(\frac{1}{4}-{m}^2\right)}{\sin^2\theta}-{\gamma} \left[\frac{b}{\sin^2\theta}+\frac{a}{\cos^2\theta}\right]\right]H(\theta)&=&0.
\label{EQ46}
\end{eqnarray}
To avoid repetition in the solution of equations (\ref{EQ46}), a careful inspection for the relationship between set of parameters $({\beta}, {\gamma})$ and the previous ones $(\widetilde{\beta}, \widetilde{\gamma})$ enable us to know that positive energy solutions for spin symmetry can be obtain directly from the pseudospin symmetric solutions by using the following parameters mapping \cite{FalThesis}:
\begin{equation}
\kappa-1\rightarrow\kappa,\ \ \tilde{\ell}\rightarrow\ell, \ \ V(r,\theta)\rightarrow-V(r,\theta),\ \ E_{n\widetilde{m}n'}\rightarrow-E_{n{m}n'} \ \ \mbox{and}\ \ C_{ps}\rightarrow-C_s.
\label{EQ47}
\end{equation}
With the help of this transformation, we arrive at the following eigensolutions for the spin symmetry
\begin{eqnarray}
\frac{E_{n{m}n'}^{(DRSK)}-M}{M+E_{n{m}n'}^{(DRSK)}-C_{s}}=-\frac{D_e^2r_e^2}{\left(n+\frac{1}{2}+\sqrt{\left(\Omega({\ell})+2n'+1\right)^2+\left(E_{n{m}n'}^{(DRSK)}+M-C_{s}\right)D_er_e^2}\right)^2}\nonumber\\
\Omega({\ell})=\sqrt{a\left(E_{n{m}n'}^{(DRSK)}+M-C_{s}\right)+\frac{1}{4}}+\sqrt{b\left(E_{n{m}n'}^{(DRSK)}+M-C_{s}\right)+{m}^2}
\label{EQ48}
\end{eqnarray}
and the corresponding eigenfunctions as
\begin{eqnarray}
f_{nmn'}^{(DRSK)}(\vec{r})&=&(-1)^{2n}\frac{\bar{N}_{nmn'}}{r\sin^{\frac{1}{2}}\theta}\frac{\Gamma(2\eta+1/2+n)}{\Gamma(2\eta+1/2)}\frac{\Gamma(2\sqrt{-{\beta}^2}+n)}{\Gamma(2\sqrt{-{\beta}^2})}\frac{exp(im\phi)}{\sqrt{2\pi}}r^\zeta e^{-\sqrt{-{\beta}^2}r}\sin^{2\eta}\theta\cos^{2p}\theta \nonumber\\
&\times&\ _1F_1\left(-n, 2\zeta, 2\sqrt{-{\beta}^2}r\right)\ _2F_1\left(-n', n'+2(\eta+p); 2\eta+\frac{1}{2}, \sin^2\theta\right),
\label{EQ49}
\end{eqnarray}
where
\begin{eqnarray} 
\zeta=\frac{1}{2}+\sqrt{\left({\ell}+\frac{1}{2}\right)^2+\gamma D_er_e},\ \ \ \eta=\frac{1}{4}\left[1+2\sqrt{{m}^2+\gamma b}\right],\ \ \ p={\frac{1}{4}\left[1+2\sqrt{\frac{1}{4}+\gamma{a}}\right]},
\label{EQ50}
\end{eqnarray}
and $\bar{N}_{nmn'}$ is the normalization constant.
\subsection{CASE 2, when $V_2(r)=\frac{kr^2}{2}$: Spin symmetry solutions of the double ring-shaped oscillator potential}
Here we consider the radial part of the problem as
\begin{equation}
\frac{d^2F(r)}{dr^2}+\left[{\beta}^2-\frac{{\gamma}kr^2}{2}-\frac{{\ell}({\ell}+1)}{r^2}\right]F(r)=0.
\label{EQ51}
\end{equation}
The angular part is still the same as equation (\ref{EQ46}). Now by using transformation (\ref{EQ47}) we can find the following solutions:
\begin{equation}
\frac{\left(M-E_{n{m}n'}^{(DRSO)}\right)\sqrt{\left(C_{s}-E_{n{m}n'}^{(DRSO)}-M\right)}}{\left(\sqrt{a\left(E_{n{m}n'}^{(DRSO)}+M-C_{s}\right)+\frac{1}{4}}+\sqrt{b\left(E_{n{m}n'}^{(DRSO)}+M-C_{s}\right)+{m}^2}+2n'+2+2n\right)}=\sqrt{2k},
\label{EQ52}
\end{equation}
and the corresponding complete wave functions as
\begin{eqnarray}
f_{n{m}n'}^{(DRS0)}(\vec{r})&=&(-1)^{2n}\frac{\bar{C}_{n{m}n'}}{r\sin^{\frac{1}{2}}\theta}\frac{\Gamma(2\eta+1/2+n)}{\Gamma(2\eta+1/2)}\frac{\Gamma\left({\ell}+n+\frac{3}{2}\right)}{\Gamma\left({\ell}+\frac{3}{2}\right)} \frac{exp(im\phi)}{\sqrt{2\pi}}r^{({\ell}+1)}e^{-\sqrt{\frac{{\gamma}k}{8}}r^2}\sin^{2\eta}\theta\cos^{2p}\theta\nonumber\\
&\times&_1F_1\left(-n, {\ell}+\frac{3}{2}, r^2\sqrt{\frac{{\gamma}k}{2}} \right)\ _2F_1\left(-n', n'+2(\eta+p); 2\eta+\frac{1}{2}, \sin^2\theta\right)
\label{EQ53}
\end{eqnarray}
where $\bar{C}_{n{m}n'}$ denotes the normalization constants.
\section{Non-relativistic limits}
Let us now study the nonrelativistic limit of our solutions by applying the following appropriate mapping $E_{nmn'}-M\rightarrow E_{nmn'}$, $E_{nmn'}+M\rightarrow\frac{2\mu}{\hbar^2}$ and with $C_s=0$, the non relativistic rovibrational energy spectrum for diatomic molecules in the presence of double ring shape Kratzer potential is given as
\begin{equation}
E_{n{m}n'}^{(NR,DRSK)}=-\frac{2\mu D_e^2r_e^2}{\hbar^2}\left[n+\frac{1}{2}+\sqrt{\left(\sqrt{\frac{2\mu a}{\hbar^2}+\frac{1}{4}}+\sqrt{\frac{2\mu b}{\hbar^2}+m^2}+2n'+1\right)^2+\frac{2\mu D_er_e^2}{\hbar^2}}\right]^{-2},
\label{EQ55}
\end{equation}
which is found identical to equation (97) of Ref. (\cite{DuY07}). The wave function can be found as:
\begin{eqnarray}
F_{n{m}n'}^{(DRSK)}(\vec{r})&=&(-1)^{2n}\frac{\bar{N}_{nmn'}}{r\sin^{\frac{1}{2}}\theta}\frac{\Gamma(2\bar{\eta}+1/2+n)}{\Gamma(2\bar{\eta}+1/2)}\frac{\Gamma(2{\bar{\beta}}+n)}{\Gamma(2{\bar{\beta}})}\frac{exp(im\phi)}{\sqrt{2\pi}}r^{\bar{\zeta}} e^{-{\bar{\beta}}r}\ _1F_1\left(-n, 2\bar{\zeta}, 2{\bar{\beta}}r\right)\nonumber\\
&\times&\sin^{2\bar{\eta}}\theta\cos^{2\bar{p}}\theta\ _2F_1\left(-n, n+2(\bar{\eta}+\bar{p}); 2\bar{\eta}+\frac{1}{2}, \sin^2\theta\right)\ \mbox{with} \ \bar{\beta}=\sqrt{\frac{2\mu E_{n{m}n'}^{(NR,DRSK)}}{\hbar^2}}\nonumber\\
\label{EQ56}
\end{eqnarray}
where
\begin{eqnarray} 
\bar{\zeta}=\frac{1}{2}+\sqrt{\left({\ell}+\frac{1}{2}\right)^2+\frac{2\mu}{\hbar^2}D_er_e},\ \ \ \bar{\eta}=\frac{1}{4}\left[1+2\sqrt{{m}^2+\frac{2\mu b}{\hbar^2}}\right],\ \ \ \bar{p}={\frac{1}{4}\left[1+2\sqrt{\frac{1}{4}+\frac{2\mu a}{\hbar^2}}\right]},
\label{EQ57}
\end{eqnarray}
and ${P}_{n{m}n'}$ denotes the normalization constant. Also, by using the same transformation, we can find the non-relativistic energy spectrum for a bound electron in the presence of a ring-shaped potential as:
\begin{equation}
E_{nmn'}^{(NR,DRSO)}=\hbar\sqrt{\frac{k}{\mu}}\left[\sqrt{\frac{2\mu a}{\hbar^2}+\frac{1}{4}}+\sqrt{\frac{2\mu b}{\hbar^2}+m^2}+2(n'+n+1)\right],
\label{EQ58}
\end{equation}
and the eigenfunctions as
\begin{eqnarray}
F_{n{m}n'}^{(DRSO)}(\vec{r})&=&(-1)^{2n}\frac{\bar{H}_{n{m}n'}}{r\sin^{\frac{1}{2}}\theta}\frac{\Gamma(2\bar{\eta}+1/2+n)}{\Gamma(2\bar{\eta}+1/2)}\frac{\Gamma\left({\ell}+n+\frac{3}{2}\right)}{\Gamma\left({\ell}+\frac{3}{2}\right)} \frac{exp(im\phi)}{\sqrt{2\pi}}r^{({\ell}+1)}e^{-\sqrt{\frac{{\mu}k}{4\hbar^2}}r}\sin^{2\bar{\eta}}\theta\cos^{2\bar{p}}\theta\nonumber\\
&\times&_1F_1\left(-n, {\ell}+\frac{3}{2}, r^2\sqrt{\frac{{\mu}k}{\hbar^2}} \right)\ _2F_1\left(-n', n'+2(\bar{\eta}+\bar{p}); 2\bar{\eta}+\frac{1}{2}, \sin^2\theta\right)
\label{EQ59}
\end{eqnarray}
where $\bar{H}_{n{m}n'}$ denotes the normalization constants. When $a=b=0$ and $2n'+\frac{1}{2}=\ell$, the solutions reduce to that of harmonic oscillator combined with the centrifugal barrier, i.e., $E_{n\ell}=\sqrt{\frac{k}{\mu}}\left(2n+\frac{3}{2}+\ell\right)$ (Example 7.3, equation (23) of ref \cite{GR1}). When $\ell=0$, the solution reduce to the one obtain recently in Ref. \cite{IkS11}
\section{Determination of the Normalization Factor}
In this section we determine the normalization constants for DRSK and DRSO wave functions. Unlike the non-relativistic case, the normalization condition for the Dirac spinor combines the two individual normalization constants $\left({N}_{n{\widetilde{m}}n'},\bar{N}_{n{m}n'}\right)$ and $\left({C}_{n{\widetilde{m}}n'},\bar{C}_{n{m}n'}\right)$ in a single integral \cite{R15}. Thus, the complete wave functions can be normalized by using the relation $\int_0^\infty\Psi\Psi^\dag dr = 1$, which explicitly implies that, for DRSK and DRSO, the normalization constant can be calculated by using
\begin{equation}
\left.\begin{matrix}
{\int_0^{2\pi}\int_0^\pi\int_0^\infty{\left(\left[g_{n\widetilde{m}n'}^{(DRSK)}(\vec{r})\right]\left[g_{n\widetilde{m}n'}^{(DRSK)}(\vec{r})\right]^\dag+\left[f_{n{m}n'}^{(DRSK)}(\vec{r})\right]\left[f_{n{m}n'}^{(DRSK)}(\vec{r})\right]^\dag\right)}}\\ 
{\int_0^{2\pi}\int_0^\pi\int_0^\infty\left(\left[g_{n\widetilde{m}n'}^{(DRSO)}(\vec{r})\right]\left[g_{n\widetilde{m}n'}^{(DRSO)}(\vec{r})\right]^\dag+\left[f_{n{m}n'}^{(DRSO)}(\vec{r})\right]\left[f_{n{m}n'}^{(DRSO)}(\vec{r})\right]^\dag\right)}
\end{matrix}\right\}r^2\sin\theta drd\theta d\phi=1.
\label{N1}
\end{equation}
Firstly, let us first obtain the normalization factor for DRSK. The upper and lower spinor components of the total wave functions can be expressed in terms of the Jacobi and Laguerre polynomials as follows:
\begin{eqnarray}
g_{n\widetilde{m}n'}^{(DRSK)}(\vec{r})&=&n!n'!)\frac{N_{n\widetilde{m}n'}}{r\sin^{\frac{1}{2}}\theta}\frac{\Gamma\left(2\sqrt{-\widetilde{\beta}^2}+n\right)}{\Gamma\left(2\sqrt{-\widetilde{\beta}^2}\right)}\frac{e^{im\phi}}{\sqrt{2\pi}}r^{\widetilde{\zeta}} e^{-\sqrt{-\widetilde{\beta}^2}r}\sin^{2\widetilde{\eta}}\theta\cos^{2\widetilde{p}}\theta\nonumber\\ 
&\times&\frac{(2\widetilde{\zeta}-1)!}{(n+2\widetilde{\zeta}-1)!}L_{n}^{2\widetilde{\zeta}-1}\left(2r\sqrt{-\widetilde{\beta}^2}\right)P_{n'}^{\left(2\widetilde{\eta}-\frac{1}{2}, 2\widetilde{p}+\frac{1}{2}\right)}\left(1-2\sin^2\theta\right),
\label{N2a}
\end{eqnarray}
\begin{eqnarray}
f_{n{m}n'}^{(DRSK)}(\vec{r})&=&n!n'!\frac{N_{n{m}n'}}{r\sin^{\frac{1}{2}}\theta}\frac{\Gamma\left(2\sqrt{-{\beta}^2}+n\right)}{\Gamma\left(2\sqrt{-{\beta}^2}\right)}\frac{e^{im\phi}}{\sqrt{2\pi}}r^{{\zeta}} e^{-\sqrt{-{\beta}^2}r}\sin^{2{\eta}}\theta\cos^{2{p}}\theta\nonumber\\ 
&\times&\frac{(2{\zeta}-1)!}{(n+2{\zeta}-1)!}L_{n}^{2{\zeta}-1}\left(2r\sqrt{-{\beta}^2}\right)P_{n'}^{\left(2{\eta}-\frac{1}{2}, 2{p}+\frac{1}{2}\right)}\left(1-2\sin^2\theta\right),
\label{N2b}
\end{eqnarray}
where we have used the relation between the hypergeometric functions, the Legendre and the Jacobi polynomials of degree $n$ in equations (\ref{N2a} and \ref{N2b}) (see Eq. (73) in \cite{YaD07} and Eq. (20) in \cite{R14})
\begin{equation}
_2F_1\left(-n, 1+\alpha+\beta+n; \alpha+1, \frac{1-z}{2}\right)=\frac{n!}{(\alpha+1)_n}P_n^{(\alpha, \beta)}(z)\ \ \ \mbox{and}\ \ \ _1F_1\left(-\gamma, m+1;z\right)=\frac{\gamma!m!}{(\gamma+m)!}L_\gamma^m(z)
\label{N3}
\end{equation}
Also, the term $(-1)^{2n}=1$. Now we by using equation (\ref{N1}), we can write
\begin{eqnarray}
&&(n!)^2(n'!)^2\frac{N^2_{n\widetilde{m}n'}}{{2\pi}}\left[\frac{\Gamma\left(2\sqrt{-\widetilde{\beta}^2}+n\right)}{\Gamma\left(2\sqrt{-\widetilde{\beta}^2}\right)}\right]^2 \int_0^\infty r^{2\widetilde{\zeta}} e^{-2\sqrt{-\widetilde{\beta}^2}r}\left[L_{n}^{2\widetilde{\zeta}-1}\left(2r\sqrt{-\widetilde{\beta}^2}\right)\right]^2dr\nonumber\\ 
&\times&\left[\frac{(2\widetilde{\zeta}-1)!}{(n+2\widetilde{\zeta}-1)!}\right]^2\int_0^\pi\sin^{4\widetilde{\eta}}\theta\cos^{4\widetilde{p}}\theta \left[P_{n'}^{\left(2\widetilde{\eta}-\frac{1}{2}, 2\widetilde{p}+\frac{1}{2}\right)}\left(1-2\sin^2\theta\right)\right]^2d\theta\int_0^{2\pi}d\phi\nonumber\\
&+&(n!)^2(n'!)^2\frac{N^2_{n{m}n'}}{{2\pi}}\left[\frac{\Gamma\left(2\sqrt{-{\beta}^2}+n\right)}{\Gamma\left(2\sqrt{-{\beta}^2}\right)}\right]^2 \int_0^\infty r^{2{\zeta}} e^{-2\sqrt{-{\beta}^2}r}\left[L_{n}^{2{\zeta}-1}\left(2r\sqrt{-{\beta}^2}\right)\right]^2dr\nonumber\\ 
&\times&\left[\frac{(2{\zeta}-1)!}{(n+2{\zeta}-1)!}\right]^2\int_0^\pi\sin^{4{\eta}}\theta\cos^{4{p}}\theta \left[P_{n'}^{\left(2{\eta}-\frac{1}{2}, 2{p}+\frac{1}{2}\right)}\left(1-2\sin^2\theta\right)\right]^2d\theta\int_0^{2\pi}d\phi=1.
\label{N4}
\end{eqnarray}
Now we transform equation (\ref{N4}) from function $(r,\theta)\longrightarrow(\xi,\varrho)$ via the following coordinate transformation $\widetilde{\xi}=2r\sqrt{-\widetilde{\beta}^2}$ or $\xi= 2r\sqrt{-{\beta}^2}$ and $\varrho=1-2\cos^2\theta$. Thus, we can find
\begin{eqnarray}
&&(n!)^2(n'!)^2\left(\frac{1}{2}\right)^{2\widetilde{p}+\widetilde{\eta}-1}\frac{N^2_{n\widetilde{m}n'}}{{4\left(2\sqrt{-{\beta}^2}\right)^{2\widetilde{\zeta}+1}}}\left[\frac{\Gamma\left(2\sqrt{-\widetilde{\beta}^2}+n\right)}{\Gamma\left(2\sqrt{-\widetilde{\beta}^2}\right)}\right]^2 \int_0^\infty \xi^{2\widetilde{\zeta}} e^{-\widetilde{\xi}}\left[L_{n}^{2\widetilde{\zeta}-1}\left(\widetilde{\xi}\right)\right]^2d\widetilde{\xi}\nonumber\\ 
&\times&\left[\frac{(2\widetilde{\zeta}-1)!}{(n+2\widetilde{\zeta}-1)!}\right]^2\int_{-1}^1(1-\varrho)^{2\widetilde{p}-\frac{1}{2}}(1+\varrho)^{2\widetilde{\eta}-\frac{1}{2}} \left[P_{n'}^{\left(2\widetilde{\eta}-\frac{1}{2}, 2\widetilde{p}+\frac{1}{2}\right)}\left(-\varrho\right)\right]^2d\varrho\nonumber\\
&+&(n!)^2(n'!)^2\left(\frac{1}{2}\right)^{2{p}+{\eta}-1}\frac{N^2_{n{m}n'}}{{4\left(2\sqrt{-{\beta}^2}\right)^{2{\zeta}+1}}}\left[\frac{\Gamma\left(2\sqrt{-{\beta}^2}+n\right)}{\Gamma\left(2\sqrt{-{\beta}^2}\right)}\right]^2 \int_0^\infty \xi^{2{\zeta}} e^{-{\xi}}\left[L_{n}^{2{\zeta}-1}\left({\xi}\right)\right]^2d\xi\nonumber\\ 
&\times&\left[\frac{(2{\zeta}-1)!}{(n+2{\zeta}-1)!}\right]^2\int_{-1}^1(1-\varrho)^{2{p}-\frac{1}{2}}(1+\varrho)^{2{\eta}-\frac{1}{2}} \left[P_{n'}^{\left(2{\eta}-\frac{1}{2}, 2{p}+\frac{1}{2}\right)}\left(-\varrho\right)\right]^2d\varrho.\nonumber
\label{N5}
\end{eqnarray}
By using the symmetry relation $P_n^{(a, b)}(x)=(-1)^nP_n^{(b, a)}(-x)$ and the following standard integral notation (see Eq. (21) in \cite{R14} and Eq. (5) of 7.391 in \cite{R16}) 
\begin{eqnarray}
&&\int_{-1}^{1}(1-x)^{a-1}(1+x)^b\left[P_n^{(a, b)}(x)\right]^2dx=\frac{2^{a+b}\Gamma(a+n+1)\Gamma(b+n+1)}{n!a\Gamma(a+b+n+1)}, \nonumber\\
&&\int_0^\infty e^{-x}x^aL_n^{a-1}(x)L_m^{a-1}(x)dx=\frac{(a+2n)\Gamma(a+n)}{n!}\delta_{nm}.
\label{N6}
\end{eqnarray}
The normalization factor becomes
\begin{eqnarray}
\left.\begin{matrix}
{N}_{n{\widetilde{m}}n'}\\ 
\bar{N}_{n{m}n'}
\end{matrix}\right\}=\frac{1}{n!n'!}\left(\begin{matrix}\left[\frac{\Gamma\left(2\sqrt{-\widetilde{\beta}^2}+n\right)}{\Gamma\left(2\sqrt{-\widetilde{\beta}^2}\right)}\right]^2\left[\frac{(2\widetilde{\zeta}-1)!}{(n+2\widetilde{\zeta}-1)!}\right]^2\left[\frac{\Gamma\left(2\widetilde{p}+n'+\frac{3}{2}\right)\Gamma\left(2\eta+n'+\frac{1}{2}\right)}{2n'!\left(2\widetilde{p}+\frac{1}{2}\right)\Gamma(2\widetilde{p}+2\widetilde{\eta}+n'+1)}\right]\left[\frac{(2\widetilde{\zeta}+2n)\Gamma(2\widetilde{\zeta}+2n)}{n!\left(2\sqrt{-{\beta}^2}\right)^{2{\zeta}+1}}\right]\\+ \left[\frac{\Gamma\left(2\sqrt{-{\beta}^2}+n\right)}{\Gamma\left(2\sqrt{-{\beta}^2}\right)}\right]^2\left[\frac{(2{\zeta}-1)!}{(n+2{\zeta}-1)!}\right]^2\left[\frac{\Gamma\left(2{p}+n'+\frac{3}{2}\right)\Gamma\left(2\eta+n'+\frac{1}{2}\right)}{2n'!\left(2{p}+\frac{1}{2}\right)\Gamma(2{p}+2{\eta}+n'+1)}\right]\left[\frac{(2{\zeta}+2n)\Gamma(2{\zeta}+2n)}{n!\left(2\sqrt{-{\beta}^2}\right)^{2{\zeta}+1}}\right]\end{matrix}\right)^{-\frac{1}{2}}
\label{N7}
\end{eqnarray}
As regard DRSO, we follow the same procedure described above and then use the following standard integral $\int_0^\infty e^{-x}x^aL_n^{a}(x)L_m^{a}(x)dx=\frac{\Gamma(a+n+1)}{n!}\delta_{nm}$ (see Eq. (3) of 7.414 in \cite{R16}), we can find
 \begin{eqnarray}
\left.\begin{matrix}
{C}_{n{\widetilde{m}}n'}\\ 
\bar{C}_{n{m}n'}
\end{matrix}\right\}=\frac{1}{(n!)^{1/2}n'!}\left(\begin{matrix}\left[\frac{\left(\widetilde{\ell}+n+\frac{1}{2}\right)!\Gamma\left(2\widetilde{p}+n'+\frac{3}{2}\right)\Gamma\left(2\eta+n'+\frac{1}{2}\right)}{4\left(-\frac{\widetilde{\gamma} k}{2}\right)^{3/4}\left(2\widetilde{p}+\frac{1}{2}\right)\Gamma(2\widetilde{p}+2\widetilde{\eta}+n'+1)}\right]\\+\left[\frac{\left(\ell+n+\frac{1}{2}\right)!\Gamma\left(2{p}+n'+\frac{3}{2}\right)\Gamma\left(2\eta+n'+\frac{1}{2}\right)}{4\left(-\frac{\gamma k}{2}\right)^{3/4}\left(2{p}+\frac{1}{2}\right)\Gamma(2{p}+2{\eta}+n'+1)}\right]\end{matrix}\right)^{-\frac{1}{2}}
\label{N7}
\end{eqnarray}
\section{Concluding Remarks}
In this paper, we have solved the relativistic Dirac equation with a certain class of non-central but separable potentials within the framework of the asymptotic iteration method. We have obtained analytical expressions for the energy levels and the corresponding normalized eigenfunctions. The comparison of our non relativistic results with the ones obtained in the literature reveals that our results are very accurate. 

Some numerical results have been presented in tables(1-4) for eigenstates of pseudospin and spin symmetry. We have utilized the following parameters $C_s=5fm^{-1}=-C_{ps}$, $r_e=0.4$, $D=15$, $M=5fm^{-1}$, and $k=1$. We have also study the effect of the ring shaped potential on the bound state energy spectrum by varying the values of $a$ and $b$, ranging from $a=b=0$, $a\neq b$, $a=b=0$. As it can be seen the existence or absence of the ring shaped potential potential has strong effects on the eigenstates of the Kratzer and oscillator  particles with a wide band spectrum except for the pseudospin-oscillator particles where there exist a narrow band gap.

It is pertinent to note that the exact results obtained here for the non-central potentials may find their applications  in the study of various mechanical systems and atomic physics.

\section*{Acknowledgments}
We thank the kind referee for the positive enlightening comments and suggestions, which have greatly helped us in making improvements to this paper. BJF thanks Prof. Dr. S. M. Ikhdair and Dr. Hamzavi for useful discussions.

\section*{Appendix A: Method of Analysis (Eigenvalues Solutions)}
\label{MET}
One of the calculational tools utilized in solving the Schr\"{o}dinger-like equation including the centrifugal barrier and/or the spin-orbit coupling term is called as the asymptotic iteration method (AIM). This interesting method was introduced in 2003 by Ciftci et al. \cite{CiE03} for finding eigenvalues problem. Ever since then, a lot of researchers have developed kin interest for this elegant approach [74-98]. For a given potential the idea is to convert the Schr\"{o}dinger-like equation to the homogenous linear second-order differential equation of the
form: 
\begin{equation}
y^{\prime\prime}(x)=\lambda_o(x)y^{\prime}(x)+s_o(x)y(x),  \tag{A1}
\label{A1}
\end{equation}
where $\lambda_o(x)$ and $s_o(x)$ have sufficiently many continous derivatives and defined in some interval which are not necessarily bounded. The differential Eq. (\ref{A1}) has a general solution \cite{CiE03, CiE05} 
\begin{equation}
y(x)=\exp\left(-\int^x\alpha(x^{\prime})dx^{\prime}\right)\left[C_2+C_1\int^x\exp\left(\int^{x^{\prime}}\left[\lambda_o(x^{\prime\prime})+2
\alpha(x^{\prime\prime})\right]dx^{\prime\prime}\right)dx^{\prime}\right]. \tag{A2}
\label{A2}
\end{equation}
If $k>0$, for sufficiently large $k$, we obtain the $\alpha(x)$ 
\begin{equation}
\frac{s_k(x)}{\lambda_k(x)}=\frac{s_{k-1}(x)}{\lambda_{k-1}(x)}=\alpha(x) \
,\ \ k=1, 2, 3.....  \label{A3} \tag{A3}
\end{equation}
where
\begin{align} 
\lambda_k(x)=\lambda^{\prime}_{k-1}(x)+s_{k-1}(x)+\lambda_o(x)
\lambda_{k-1}(x),\nonumber\\
s_k(x)=s^{\prime}_{k-1}(x)+s_o(x)\lambda_{k-1}(x) \ ,\ \ k=1, 2, 3..... \tag{A4}
\label{A4}
\end{align}
with quantization condition 
\begin{equation}
\delta_k(x)= \left| 
\begin{array}{lr}
\lambda_k(x) & s_k(x) \\ 
\lambda_{k-1}(x) & s_{k-1}(x)
\end{array}
\right|=0\ \ ,\ \ \ k=1, 2, 3....  \tag{A5}
\label{A5}
\end{equation}
For a given potential, the idea is to convert the radial Schr\"{o}dinger-like equation to the form of equation (\ref{A2}). Then $\lambda_0(x)$ and $s_0(x)$ are determine and $s_k(x)$ and $\lambda_k(x)$ parameters are calculated by the recurrence relations given by equation (\ref{A4}). The energy eigenvalues are then obtained by the condition given by equation (\ref{A5}) if the problem is exactly solvable. For nontrivial potentials that have no exact solutions, for a specific $n$ principal quantum number, we choose a suitable $x_0$ point, determined generally as the maximum value of the asymptotic wave function or the minimum value of the potential [96-100] and the approximate energy eigenvalues are obtained from the roots of equation (\ref{A5}) for sufficiently great values of $k$ with iteration for which $k$ is always greater than $n$ in the numerical solutions.

\section*{Appendix B: Method of Analysis (eigenfunction Solutions)}
\label{EEFUN}
Suppose we wish to solve the radial Schr\"{o}dinger-like equation for which the homogenous linear second-order differential equation takes the following general form
\begin{equation}
y''(x)=2\left(\frac{ax^{N+1}}{1-bx^{N+2}}-\frac{m+1}{x}\right)y'(x)-\frac{Wx^N}{1-bx^{N+2}}y(x). \tag{B1}
\label{A6}
\end{equation}
The exact solution $y_n(x)$ can be expressed as \cite{CiE03, CiE05}
\begin{equation}
y_n(x)=(-1)^nC_2(N+2)^n(\sigma)_{_n}{_2F_1(-n,\rho+n;\sigma;bx^{N+2})}, \tag{B2}
\label{A7}
\end{equation}
where the following notations has been used
\begin{equation}
(\sigma)_{_n}=\frac{\Gamma{(\sigma+n)}}{\Gamma{(\sigma)}}\ \ ,\ \ \sigma=\frac{2m+N+3}{N+2}\ \ \mbox{and}\ \ \rho=\frac{(2m+1)b+2a}{(N+2)b}. \tag{B3}
\label{A8}
\end{equation}

\end{document}